\documentclass[9pt,twocolumn,twoside]{osajnl}
\journal{ol} 

\usepackage[squaren,Gray,cdot]{SIunits}
\usepackage{subcaption}
\usepackage{xcolor}
\usepackage{stfloats}
\usepackage{float}

\newcommand{\fundingfont}{\normalfont\sffamily\bfseries\fontsize{9}{10}\selectfont}
\setboolean{shortarticle}{true}

\title{Multimodal imaging for intra-droplet gas-cavity observation during droplet fragmentation}

\author[1]{Luc Biasiori-Poulanges}
\author[1*]{Hazem El-Rabii}

\affil[1]{Institut Pprime, CNRS UPR 3346 -- Universit\'e de Poitiers -- ISAE-ENSMA, 1 avenue Cl\'ement Ader, 86961 Futuroscope, France}

\affil[*]{Corresponding author: hazem.elrabii@cnrs.pprime.fr}

\dates{Compiled \today}

\begin{abstract}
Herein, planar laser-induced fluorescence (PLIF) is used in combination with shadowgraphy to study water-droplet aerobreakup. The acquired shadowgraph data are in agreement with previous visualization studies but differ from the PLIF results, yielding new insights into the fragmentation process. In particular, the PLIF data reveal changes in droplet topology during fragmentation that result from the entrapment or formation of gas cavities inside the liquid phase. In some instances, topological modification can be observed to arise from the presence of these cavities. In addition, the cavities may act as weak spots, facilitating droplet split-off.
\end{abstract}

\setboolean{displaycopyright}{true}

\begin{document}

\maketitle

Droplet fragmentation subsequent to shock-wave impact is a critical event in a host of natural and engineering applications \cite{gelfand1996,joseph1999, guildenbecher2017characterization}. As such, this process has been discussed extensively in the literature, and thus several breakup regimes have been identified on the basis of fluid properties and flow parameters. The prevailing view is that the droplet breakup modes can be classified into five regimes, namely, the vibrational, bag, bag and stamen, stripping, and catastrophic breakup regimes. A qualitative description of these regimes can be found in \cite{guildenbecher2009}. Droplet deformation and breakup patterns have traditionally been examined using direct visualization and shadow-based techniques. Useful information -- for example, identification and clarification of the physical mechanisms behind the breakup regimes -- have been gained via the use of these techniques. However, they suffer from the serious drawback of relying integration along the light path, which severely limits their applicability for the study of aerobreakup processes that are inherently three-dimensional \cite{jalaal2014transient}. In addition, some regimes are characterized by the presence of a mist surrounding the core droplet structure. This mist acts as a curtain, obstructing the view of the bulk liquid and hindering direct observation. Finally, limited image contrast renders the use of shadowgraphy inappropriate for attempting to distinguish liquid from vapor phases and may therefore present a biased view of fragmentation processes \cite{biasiori-poulanges2019}.

In this Letter, we examine two key situations in which these shortcomings represent severe impediments such that  reliance on shadowgraphy alone may result in misleading conclusions. Specifically, the stripping breakup regime and shock interaction with a drop of liquid containing a bubble are investigated. Our interest in these two droplet systems  is motivated as follows. In the first case, it is known that a gas cavity can be generated inside a droplet under the action of an intense shock wave \cite{sembian2016}. The presence of such bubble, which causes major changes in interfacial dynamics \cite{bhattacharya2016}, may lead to a new form of fragmentation. (We refer the reader to \cite{liang2020} for a very recent investigation related to this topic.) Regarding the second case, the droplet-bubble ensemble represents an interesting case study to investigate Rayleigh-Taylor-type curved interface instabilities. It should be noted that, in both cases, the early stages of drop-shock interaction are an inherent part of the breakup process, which means that the use of shadowgraphy, to follow the interaction, remains essential. Therefore, the strategy adopted here is to combine shadowgraphy with fluorescence imaging, to compensate for the aforementioned drawbacks of the former technique. In particular, an induced-fluorescence technique in light-sheet configuration is applied to yield two-dimensional time-resolved measurements. Crucially, the fluorescence imaging technique has the advantage of a large disparity in the fluorescence signal intensities for the two phases (vapor and liquid), enabling the phases to be distinguished. 

Fluorescence imaging, as implemented in our study, involves injecting a fluorescent substance into a fluid and detecting its presence by laser-induced fluorescence measurement. As an alternative to the commonly used Rhodamine dyes (i.e., Rhodamine B and 6G), we chose eosin yellowish (EY), a brominated derivative of fluorescein, as a tracer substance. The selection of EY was motivated by several considerations. First, EY is a water-soluble dye (300~\gram\per\liter) that has a fluorescence quantum yield in water of 0.36 \cite{zhang2014}. This quantum yield is sufficiently high to achieve detectable fluorescence signals at low dye concentrations. Second, the emission spectrum of EY exhibits a reasonably large Stokes shift, which enables the excitation wavelength to be filtered from the emission signal, without significant loss of signal fluorescence intensity. Finally, EY has low toxicity. In this study, we used a pharmaceutically available aqueous solution of EY (Cooper S. A.) that is typically used as an adjunctive treatment for skin lesions. The composition of the solution, according to the manufacturer, was 2 \gram~of EY per 100 \milli\liter~of purified water, plus negligible, but unspecified, amounts of chlorphenesin and pentylene glycol. This solution was used as base for preparing water tagged with EY; the pharmaceutical solution was added dropwise to 100 \milli\liter~of distilled water. We found that the minimum amount of EY required to produce an acceptable fluorescence signal-to-noise ratio (SNR) was 200~\milli\gram\per\liter. The effect of EY concentration on the viscosity and surface tension of distilled water was also addressed \cite{biasiori2020data}. These two physical quantities were measured for distilled water, with and without EY, using a rotational rheometer (D-HR2, TA Instruments). The addition of 1 \gram~of EY per liter of water was found to cause a change in the dynamic viscosity of up to 5\%, over temperatures up to 80\degree C. Surface tension, which was measured at room temperature using a drop shape analyzer (DSA25, Kr\"uss GmbH), was found to be unaffected by the addition of EY ($\sim0.5\%$). 
\begin{figure}[htbp]
	\begin{center}
		\includegraphics[width=\columnwidth]{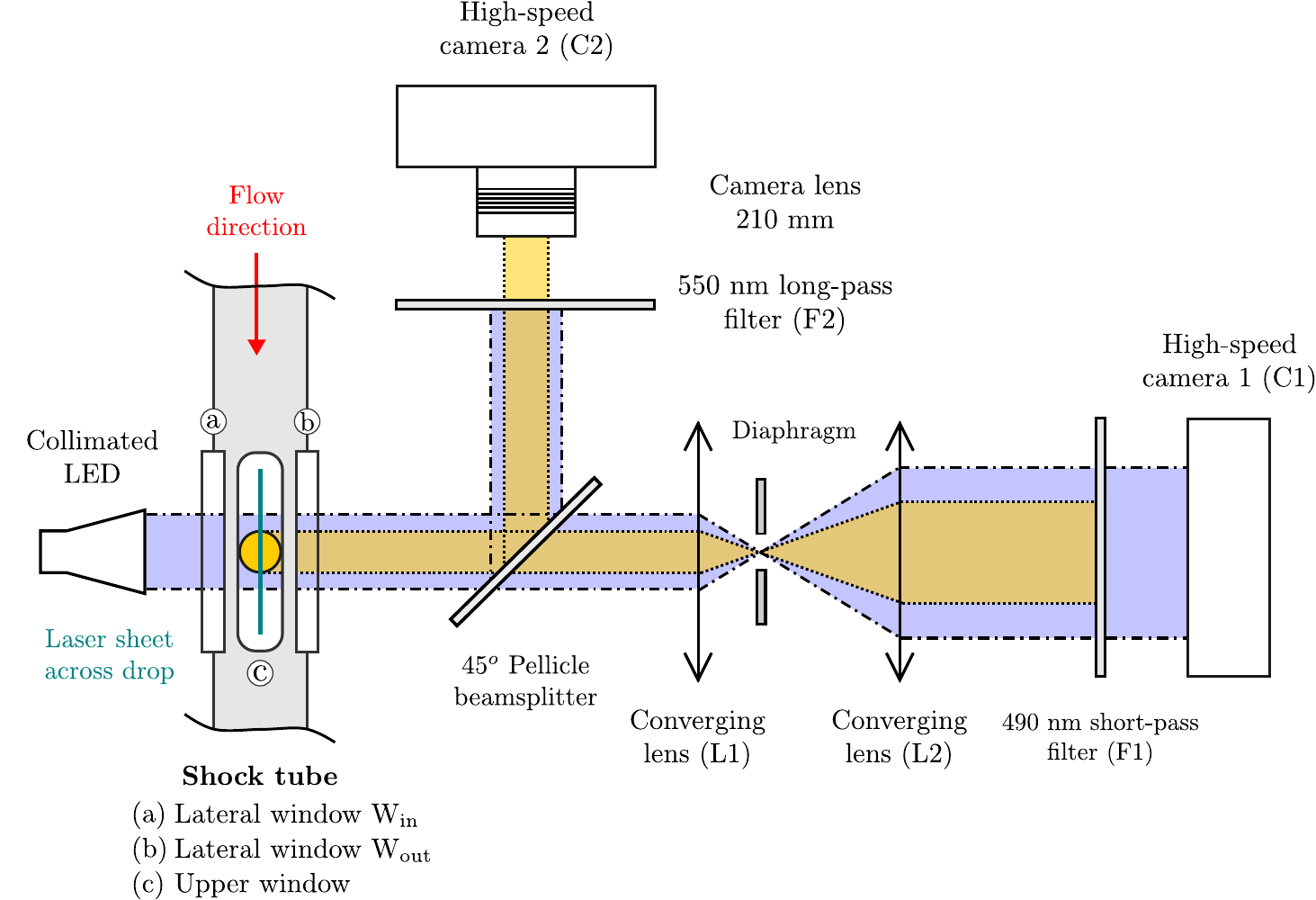}
		\caption{Sketch of the experimental arrangement (top view).}
		\label{figure1}
	\end{center}
\end{figure}

To produce a transient flow of known velocity, the experiments were conducted in an air shock tube with a double-membrane arrangement. The driven section of the tube comprises a test chamber with a square cross-section. To allow optical access, the sidewalls of the test chamber were fitted with two oblong BK7 windows that span its entire height and length. An additional thin window was installed in the top wall of the test chamber to provide laser access. During the experiments, an EY dye-doped water drop was held in a stable equilibrium at the center of the test chamber by the acoustic radiation pressure of an ultrasonic standing wave generated by a single-axis acoustic levitator. The levitation system was a Langevin-type transducer coupled to a mechanical amplifier; the latter had a radiating surface mounted flush with the inside lower surface of the chamber. The upper surface of the chamber acted, therefore, as an acoustic wave reflector for standing-wave generation. The optical setup adopted for the acquisition of simultaneous shadowgraph and PLIF images is shown in Fig.~\ref{figure1}. High-speed shadowgraphs of aerobreakup events were captured by a high-speed camera (C$_1$; Fastcam Photron SA-Z) at 96,000 frames per second, using a 1-\micro\second~exposure time. A light-emitting diode (LED), combined with telecentric optics (Opto Engineering S.r.l.), emitting a collimated beam parallel to the optical axis is used as an illumination source to backlight the drop. The 1-\watt~LED produces a uniform and continuous blue light with an optical bandpass (FWHM) of 60~\nano\meter~centered at 460~\nano\meter. The light beam emitted by the LED is 60 mm in diameter and is transmitted across the test section through its side windows (W$_\mathrm{in}$, W$_\mathrm{out}$). Shadow images of the drop are projected onto the 1024$\times$1024-pixel sensor of the camera (C$_1$) by an afocal pair of lenses, L$_1$ ($f=1000~\milli\meter$) and L$_2$ ($f=400~\milli\meter$). A 6-\milli\meter-diameter iris diaphragm is placed between L$_1$ and L$_2$ to limit the effective aperture of the system. A short-pass filter (F$_1$) is set in front of the camera sensor to suppress light wavelengths above 490~\nano\meter. Simultaneous PLIF imaging from the same perspective is accomplished by positioning a 45\degree~pellicle beamsplitter between W$_\mathrm{out}$ and L$_1$. The beamsplitter produces two identical images by reflecting 50\% of the image-forming light and transmitting the remainder. The EY fluorescence is excited by a continuous wave 532-\nano\meter~laser (a 5-\watt~frequency doubled Nd:YVO$_4$ laser, Verdi V5, Coherent Inc.). Prior to entering the test chamber from the upper window, the laser beam is shaped into a laser sheet (200 \micro\meter~thick) by diverging the beam in the direction of the tube axis using a cylindrical lens ($f=-10~\milli\meter$), focusing the beam in the perpendicular direction with a 200 \milli\meter~spherical lens. The laser sheet intersects the drop at its median plane. The emitted fluorescence radiation reflected by the beamsplitter is imaged onto a second high-speed camera C2 (Fastcam Photron SA-Z) via a 105-\milli\meter~$f/2.8$ macro lens (Sigma Corp.) coupled to a $2\times$ teleconverter. The camera is operated with a pixel resolution of 512$\times$344 and the same time settings as for C$_1$. A long-pass filter (F$_2$) in front of the camera optics blocks light of wavelengths less than 550~\nano\meter. The two cameras are simultaneously triggered by a pressure signal generated by the shock wave.

Under continuous excitation, the fluorescence signal was observed to decrease over time. This continuous fading of the fluorescence signal is attributed to a photochemical modification of EY resulting in the irreversible loss of its ability to fluoresce (photobleaching). This phenomenon severely reduces the SNR of the detected signal, and thus the observation time window is limited. The fluorescence signal loss caused by photobleaching was quantitatively determined by measuring the gradual loss in fluorescence intensity in a time-lapse image series of the same field of view. To establish the extent of the signal loss, all the images were normalized to the first image taken. Investigations into the temporal evolution of the fluorescence signal generated under 2.5-\watt~laser excitation of a levitated droplet revealed that photobleaching took place on timescales on the order of milliseconds, with the fluorescence signal decreasing exponentially. Specifically, we observed that the fluorescence intensity decreased by 20\% within a few milliseconds. This time scale, however, is sufficiently long to allow fluorescence imaging in droplet aerobreakup studies. To minimize EY photobleaching, the drop was exposed only to the excitation light during image acquisition. 

\begin{figure}[H]
	\begin{center}
		\centering
		\includegraphics[width=\columnwidth]{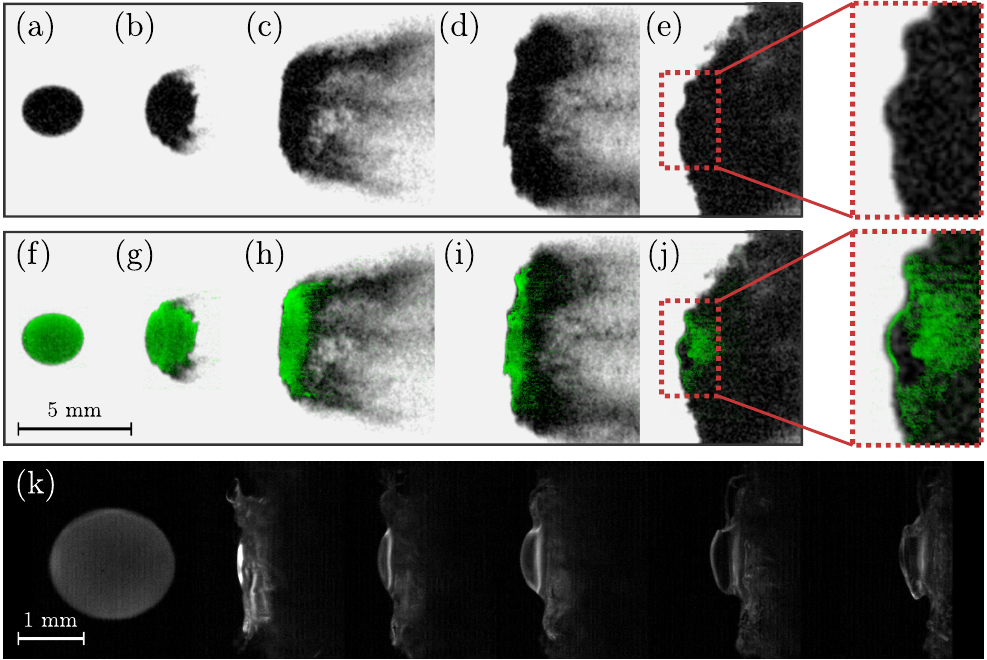}
		\caption{Images of a spherical gas cavity during aerobreakup ($\mathrm{We}=1447$). Times, in \micro\second: (a,f) $<$0, (b,g) 129, (c,h) 289, (d,i) 419 and (e,j) 779. (k) Image sequence showing the formation of another gas cavity during aerobreakup with a frame interval of  25~\micro\second.}
		\label{figure2}
	\end{center}
\end{figure}

\begin{figure*}[b]
	\begin{center}
		\centering
		\includegraphics[width=.99\textwidth]{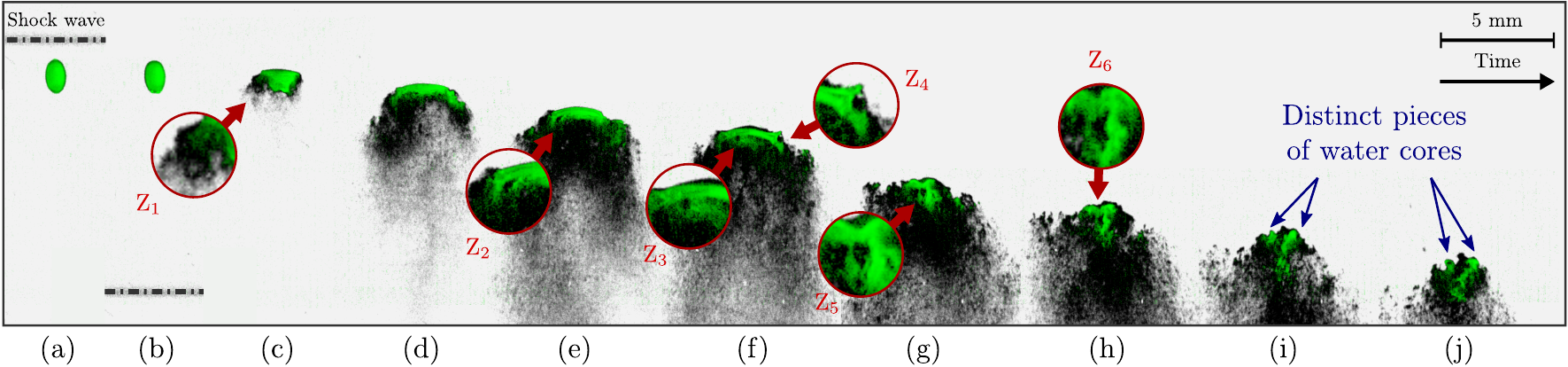}
		\caption{Appearance of a gas cavity during aerobreakup ($\mathrm{We}=762$). Times, in \micro\second: (a) $<$0, (b) 25, (c) 98, (d) 202, (e) 307, (f) 359, (g) 463, (h) 505, (i) 546 and (j) 588.}
		\label{figure3}
	\end{center}
\end{figure*}

\begin{figure}[htbp]
	\begin{center}
		\centering
		\includegraphics[width=\columnwidth]{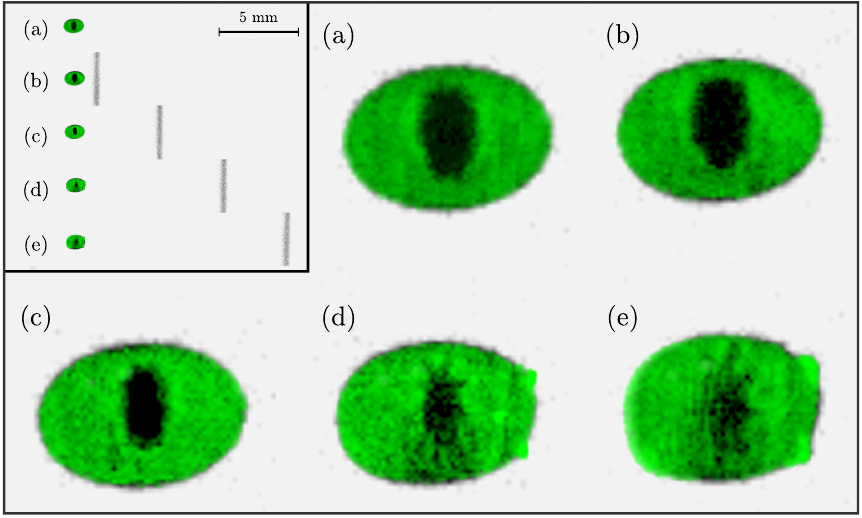}
		\caption{Interaction of a planar shock wave with a water droplet having an initially embedded air bubble ($\mathrm{We}=701$). Times, in \micro\second: (a) <0, (b) 5, (c) 15, (d) 26 and (e) 36.}
		\label{figure4}
	\end{center}
\end{figure}

To assess the ability of the PLIF technique to observe gas cavity formation and evolution within a water droplet undergoing fragmentation, we conducted two types of experiments. In both cases, fragmentation occurred in the flow behind a Mach 1.3 shock wave, i.e., in an air flow of 155~\meter\per\second. In the first experiment, we considered the interaction of the shock wave with a bubble-free droplet. Two different droplet diameters, corresponding to Weber numbers (ratio between inertia and surface tension) of $\mathrm{We}=762$ and $\mathrm{We}=1447$, were examined in this experiment. The second experiment involved a water droplet with an embedded gas cavity existing prior to the shock interaction. For this configuration, the Weber number, based on the liquid droplet diameter, was 701. The uncertainty on the Weber numbers was $\pm 3\%$. In all the images shown in this Letter [Figs.~\ref{figure2}-\ref{figure4}, except Fig.\,\ref{figure2}(k)], fluorescence intensity is represented in green, superimposed on the simultaneously acquired grayscale shadowgraph image. 

Figure~\ref{figure2} shows an example image sequence captured after the interaction of a levitated oblate drop with the shock wave, for $\mathrm{We}=1447.$ The shock wave direction is from left to right. The upper row of images displays shadowgraphs of the fragmentation process, and the middle row shows the same sequence with fluorescence intensities superimposed. The shadowgraph sequence shows that in the very first moments following the interaction, the water droplet displays no response to the passage of the shock wave [Fig.~\ref{figure2}(a)]. The droplet appears to be quiescent for about 30~\micro\second, before it starts to undergo deformation. At the end of this period, the effects of the interaction are manifested by the appearance of small lips, on the top and bottom of the leeward face of the drop. The formation of the lips is immediately followed by the development of ripples over the entire surface of the drop. As time elapses, the ripples and lips increase in magnitude and bend toward the flow direction, producing small trails of mist under the destabilizing effect of the airstream. The drop simultaneously undergoes a gradual flattening in the streamwise direction, with the windward face maintaining a nearly spherical cap shape and the leeward face being flattened into a planar surface. First, the deformation results in muffin-like shaped droplet without any noticeably change in its lateral dimension [Fig.\,\ref{figure2}(b)]. In a second stage, during which the flattening process proceeds, the extent of the droplet increases significantly (up to twice the initial value) in all directions perpendicular to the air flow direction, while the mist grows in volume and density [Fig.\,\ref{figure2}(c)]. It is worth noting that the mist extends out from the water droplet in a brush-like shape. The appearance of such a mist structure is very likely to be the result of a combination of instabilities, as suggested in \cite{jalaal2014transient}. Beside the mist flow pattern, the flattening of the entire windward face of the droplet, except for the edges that bend in the air flow direction, should be noted. It is also noteworthy that the drop, which is essentially a flat disc at this point in time, deforms into a nearly cowboy-hat drop as time progresses [Fig.\,\ref{figure2}(d)]. In the final frame [Fig.\,\ref{figure2}(e)], the droplet and the mist fill a downstream portion of the entire image space, being seen in the shadowgraph as a dark silhouette with large corrugations on its windward face. Figure\,\ref{figure2}(f)-(i) shows the fragmentation progression, as imaged by both techniques.

We note that the first discrepancy between the techniques is apparent in Fig.\,\ref{figure2}(g), where the mist flow is not captured by the fluorescence measurement. This difference can be understood insofar as the mist is made up of gas and droplets that are too fine to produce detectable fluorescence signals. Next, the droplet core seen in Fig.\,\ref{figure2}(h) presents as a quasi-crescent-shaped body with small horns, rather than a flat disc. The fact that the horns are seen in the fluorescence image indicates that they are water ligaments. These extend beyond the leeward face and produce the trails of mist from their tips. We further note that there is no evidence for the presence of other ligaments in the water volume intercepted by the laser sheet. This suggests that the other trails of mist seen in the shadowgraphs are emitted from the periphery of the drop, and therefore the water core can be visualized as a jellyfish-shaped body. In the next frame, Fig.\,\ref{figure2}(i), the fluorescence image shows that the windward face of the droplet has a cowboy-hat shape, in perfect agreement with the shadowgraphy data. The fluorescence measurements allow further specification of the actual droplet body shape, which in fact is far less thick than had been assumed from the shadowgraph image. The final frame of the sequence, Fig.\,\ref{figure2}(j), strikingly illustrates that sole reliance on shadowgraph visualization may lead to incorrect conclusions on the process of aerobreakup. What appears to be an intense deformation of the windward surface in the shadowgraph [Fig.\,\ref{figure2}(e)] may be misinterpreted as resulting from the development of instabilities (see \cite{joseph1999}, for instance). Fluorescence imaging gives quite a different depiction of the origin of this deformation: the bump emerging on the windward face results from the growth of a gas cavity inside the liquid core. This gas cavity is trapped between a spherical liquid film (upstream) and the remaining liquid mass (downstream). Ultimately, the spherical film, whose thickness is about 100~\micro\meter, bursts open in the opposite-flow direction (not shown in Fig.~\ref{figure2} ). We point out that repeated experiments conducted in the same conditions confirmed the systematic occurrence of this gas cavity. Another example of such an occurrence being captured by a high-resolution PLIF arrangement is displayed (in grayscale) in Fig.\,\ref{figure2}(k). The latter sequence was recorded using a frame interval of 25~\micro\second.

Figure~\ref{figure3} shows sequential images of different stages in the fragmentation of a levitated oblate drop, for $\mathrm{We}=762.$ The shock wave direction is from top to bottom. We note, as in the previous example, that both techniques reveal a similar picture of the first part of the breakup process [Fig.\,\ref{figure3}(a)–(d)]. Unlike shadowgraphy, which images the entire droplet, PLIF  provides spatial resolution within the droplet, revealing illuminating features of the fragmentation process. In this respect, our experiments highlight changes in the droplet topology that shadowgraphy is not able to detect. These changes start to be observable in the fluorescence intensity at $t= 307~\micro\second$. Indeed, inspection of the expanded views of the image in Fig.\,\ref{figure3}(f) reveals two dark spots on the green background (Z$_{3}$ and Z$_{4}$). These non-fluorescing regions are evidence of the presence of gas cavities inside the liquid phase. Their proximity to the droplet surface suggests that they may result from the entrapment of air through the droplet surface. A plausible scenario for the mechanism of entrapment is that crests or surface deformations entrap pockets of air during folding onto the bulk liquid (Z$_2$ and Z$_4$ may show a snapshot of this entrapment process). In Fig.\,\ref{figure3}(g), the shadowgraph shows the fragmenting droplet as a single dark silhouette with some sagging of the windward face. This sagging may indicate that the remaining part of the drop is no longer a coherent structure and may possibly consist of mist. The fluorescence image, however, shows that at least a core of water remains. A portion of liquid water is embedded with an air pocket, as evidenced by the dark area visible inside the fluorescence zone (Z$_5$). As time elapses, the air pocket is distorted and stretched in the streamwise direction (Z$_6$) until its boundary reaches the droplet surface, and ultimately shatters the droplet into two distinct cores, which are indicated by the blue arrows in Fig.\,\ref{figure3}(i)-(j). These observations demonstrate that gas cavities may act as weak spots at which the liquid is subsequently torn apart. 

Lastly, we considered the interaction of a shock wave with a compound droplet, which consists in an oblate water droplet containing a single air bubble ($\mathrm{We}=701$). The bubble was inserted into the droplet by means of a millimetric syringe filled with air. The sequence of images in Fig.\,\ref{figure4} shows the evolution of the compound droplet after it was hit by the shock, over a period of  36~\micro\second. Both droplet and bubble are in a stable equilibrium prior to the interaction. In the early stages following the interaction, the increased pressure due to the shock passage causes the droplet to contract and water to be set in motion toward the interior of the droplet. Under the action of liquid pressure, the bubble undergoes a rapid contraction [Fig.\,\ref{figure4}(b)-(c)]. This contraction is immediately followed by the development of small ripples at the air bubble–water interface, which can be clearly seen in the expanded view of in  Fig.\,\ref{figure4} (d). The ripples then grow quickly in response to the interface acceleration, until they reach the droplet surface [Fig.\,\ref{figure4}(e)]. In Fig.\,\ref{figure4}(e), the bubble is seen to exhibit a star-like shape with eight arms. To the best of our knowledge, this is the first observation of such a pattern in the context of the interaction of a shock wave with a water droplet initially embedded with a gas bubble. 

To conclude, we have demonstrated that PLIF provides additional spatially-resolved details on the process of water droplet fragmentation, illuminating aspects that shadowgraphy could not detect. The complementarity of the techniques was illustrated in three different experiments, which provided new insights into the aerobreakup process. In particular, the results presented in this Letter reveal changes in droplet topology that result from the entrapment or formation of a gas cavity inside the liquid phase during fragmentation. In some instances, topological modification was observed to arise from the presence of these cavities, which may act as weak spots for droplet split-off. Finally, our experiments also show that the combination of PLIF and shadowgraphy is specifically useful for the investigation of the aerobreakup of water droplets with initially embedded gas cavities. 

~\\
\noindent {\fundingfont{Funding.}} R\'egion Nouvelle-Aquitaine (SEIGLE project 2017-1R50115; CPER FEDER project). \\

\noindent {\fundingfont{Disclosures.}} The authors declare no conflicts of interest.

\bibliography{references}

\end{document}